\documentstyle[12pt,epsfig]{article}

\newskip\humongous \humongous=0pt plus 1000pt minus 1000pt
  \newif\ifdtup

\def\frac#1#2{ {{#1} \over {#2} }}

\def\third{\mbox{\small $\frac{1}{3}$}}

\def\VEV#1{\left\langle #1\right\rangle}

\def\ie{\hbox{\rm i.e. }}

\def\beq{\begin{equation}}
\def\eeq{\end{equation}}

\def\non{\nonumber}
\def\beqn{\begin{eqnarray}}
\def\eeqn{\end{eqnarray}}

\def\L{\Lambda}
\def\be{\beta}

\def\g{\gamma}
\def\as{\alpha_{\sf s}}

\def\Tr{\mbox{Tr}\;}
\def\MSbar{\overline{\rm MS}}

\def\Journal#1#2#3#4{{#1} {\bf #2}, #3 (#4)}
\def\NPB{{\em Nucl. Phys.} B}
\def\PLB{{\em Phys. Lett.}  B}

\def\PR{\em Physics Reports}

\begin{document}
\begin{titlepage}
\begin{flushright}
     UPRF-2000-16\\
     SWAT: 281\\
     November 2000 \\
\end{flushright}
\par \vskip 10mm
\begin{center}
{\Large \bf A consistency check for Renormalons in Lattice Gauge Theory: 
$\beta^{-10}$ contributions to the $SU(3)$ plaquette\footnote{Research 
supported by Italian MURST under contract 9702213582, 
by I.N.F.N. under {\sl i.s. PR11} and by EU Grant EBR-FMRX-CT97-0122.}}
\end{center}
\par \vskip 2mm
\begin{center}
F.\ Di Renzo$\,^a$,
and  L.\ Scorzato$\,^{a,b}$ \\
\vskip 5 mm
$^a\,${\it Dipartimento di Fisica, Universit\`a di Parma \\
and INFN, Gruppo Collegato di Parma, Italy}
\vskip 2 mm
$^b\,${\it Department of Physics, \\
University of Wales at Swansea, United Kingdom}\\
\end{center}
\par \vskip 2mm
\begin{center} {\large \bf Abstract} \end{center}
\begin{quote}
We compute the perturbative expansion of the Lattice $SU(3)$ plaquette 
to $\beta^{-10}$ order. The result is found to be consistent both with the 
expected renormalon behaviour and with finite size effects on top of that. 
\end{quote}
\end{titlepage}

\section{Introduction}

In recent years, the Numeric implementation of Stochastic Perturbation 
Theory (NSPT) was introduced, which was able to reach 
unprecedented high orders in pertubative expansions in Lattice Gauge 
Theory (LGT). The historic success of NSPT was the computation of the Lattice 
$SU(3)$ basic plaquette to order $\beta^{-8}$\cite{8loop}. This led 
to the possibility of actually verifying the expected dominance of the 
leading InfraRed (IR) Renormalon \cite{beneke} associated to a dimension 
four condensate. The computation was performed on a $8^4$ 
lattice. We are now in a position to quote results to a reasonable accuracy 
for the perturbative expansion of the same quantity to order 
$\beta^{-10}$ on both a $8^4$ and a $24^4$ lattice. \\
There are at least three main motivations for such a computation. First 
of all, it provides a clear consistency argument in support of the 
conclusions of our previous work. The results in \cite{8loop} are actually 
the only ones from which one can recognize the presence of 
renormalons by direct inspection of the perturbative coefficients of a QCD 
quantity. Ironically, this is true in a scheme (the lattice) which 
one would never choose for a high order QCD calculation in a standard 
(as opposed to NSPT) approach. We will give strong evidences that a 
very good control has been taken over both the IR renormalon growth and 
the finite size effects on top of that. In particular, finite size effects 
have always been a matter of concern for the computation. They play a 
fundamental role since they impinge on the IR region which is responsible 
for the renormalon growth. Basically, the higher is the loop, 
the lower are the (IR) momenta which give the main contribution. 
In view of this one could be 
worried that one is missing important contributions on the $8^4$ lattice on 
which the NSPT perturbative computations were first performed on. 
In \cite{firstFV} evidences from the non linear sigma models were collected 
which suggested that renormalon growth was not substantially tamed at the 
eight loop level. We can now strengthen our confidence that the leading 
behaviour had been actually properly singled out in our previous work. 
On top of that we are now in a position to assess the finite size effects, 
which turn out to be of the order of a few percent at the order we got. \\
While the eight loop expansion in \cite{8loop} was a major success, it 
actually opened more problems than it settled down and this sets the stage 
for the second motivation for our present computation. There was a 
longstanding problem concerning the possibility of determining the Gluon 
Condensate (GC) from LGT by a method which actually 
relied on the computation of high orders in the 
perturbative expansion of the plaquette. The latter is the lattice 
representation for the GC and its perturbative expansion is an additive 
renormalisation which is by far the leading contribution at the value of 
$\beta$ usually taken into account. This perturbative contribution can be 
seen as the contribution coming from the identity operator in the Operator 
Product Expansion (OPE) for the plaquette. The key point is that in this 
OPE there is no space for an order $a^2$ term, a gauge invariant operator of 
dimension two missing. Next term in OPE is expected to be an $a^4$ 
contribution associated with the genuine GC. 
The old approach to the problem was to subtract the first (perturbative) 
contribution from Monte Carlo measurements of the plaquette in order to 
single out the $a^4$ term (the GC) \cite{pisa}. The signature of such a 
term can be recognized by performing the subtraction at different values 
of $\beta$ and looking for the scaling dictated by Asymptotic Freedom. 
The actual problem is that the series for the plaquette ``not only has 
to be subtracted, it has to be defined in first place''\footnote{This is a 
citation from Beneke's review \cite{beneke}}. 
Due to the presence of the IR renormalon, the series 
has an ambiguity in different prescriptions for resumming it which is 
just the same order as the GC, {\em i.e.} an $a^4$ contribution. Being aware 
of this, the most striking point is that by performing the subtraction, with 
any given prescription, one is left with an $a^2$ contribution, at least 
on the $8^4$ lattice the computation was first performed on \cite{L2}. The 
problem challenges our understanding of potentially any short-distance 
expansion and so it needs to be cleared up. At the moment more than one 
explanation appear to be possible. 
The effect could either be a pure lattice artifact or a fundamental issue. 
With respect to the latter hypothesis, \cite{L2} suggested the effect of  
UltraViolet (UV) contributions to the running coupling. 
Since the time of \cite{L2} the phenomenology of non--standard $Q^{-2}$ 
power effects has actually been reported growing. In particular, 
the work in \cite{ZakCGP} establishes a relation between a dual 
superconductor confinement mechanism and ``OPE--violating'' corrections. 
In this paper we do not want to enter at all the {\em $a^2$--affair}. 
Still, the reliability of our previous perturbative computations has to 
be assessed with this question in mind. Given the potential impact of the 
result in \cite{L2}, we want to be sure that nothing was wrong with respect 
to the Renormalon contribution extraction, leaving for a subsequent 
publication \cite{testL2} a refinement of the whole analysis concerning 
the subtraction procedure. We stress that such a refinement will in 
particular benefit by the study of finite size effects.\\
Finally something can be learned from the computation about the NSPT method 
itself. The present result establishes the highest order ever reached by 
NSPT in LGT. In \cite{toymodels} a word of caution was 
spent on uncritical application of the method, stressing the issue of 
a careful assessment of statistical errors. Ironically enough, it was 
pointed out that problems plague simple models much more than complex 
systems like Lattice $SU(3)$. We refer the reader to a future 
publication \cite{lgtNSPT} for a more organic report on the application 
of the NSPT method to LGT. Still, it is reassuring 
(although of course admittedly not a proof of reliability of the method) 
to note that in the context of this work there is a fine agreement 
between the results obtained by our stochastic method and those 
inferred by a (maybe strong) theoretical prejudice. \\
The paper is organized as follows. In Sect.~2 we recall the basics 
of IR Renormalon analysis applied to a $dim = 4$ condensate and write 
the formulas relevant for a finite lattice, which are the foundations of 
our subsequent analysis. In Sect.~3 we present our 
results for the first ten coefficients in the expansion of the plaquette
both on a $8^4$ and on a $24^4$ lattice, showing that they are consistent 
with the Renormalon factorial growth and with the expected finite size 
effects on top of that. Sect.~4 contains our conclusions and perspectives 
for future work. An appendix covers algebraic details.

\section{The IR Renormalon on a finite lattice}

We start by writing first of all the definition of our observable, that 
is the basic plaquette normalised as 
\begin{equation}
W(M) \, = \, 1 - \third \VEV{\Tr U_p} \,,
\end{equation}
$U_p$ being the product of links around a $1\times1$ Wilson Loop. 
We write an explicit dependence on the size ($M$) of the lattice on which 
the osbervable is computed. In Perturbation Theory 
\beq\label{eq:clat}
W(M) = \sum_{\ell=1} c^{(M)}_\ell \be^{-\ell} 
\eeq
Next we move to recall the main points of Renormalon analysis. 
Notations are slightly different from those of \cite{8loop,L2} and closer 
to those of \cite{firstFV}. 
The aim of this section is anyway to be self--contained. 

\subsection{Basics on the factorial growth of perturbative coefficients}
The expected form for a dimension $4$, Renormalisation Group invariant 
condensate is written as
\beq\label{eq:Int1}
W =\int^{Q^2}_0\; \frac{k^2\,dk^2}{Q^4} \; f(k^2/\L^2) \,.
\eeq
$Q$ is in our case the UV cutoff fixed by the lattice spacing: $Q=\pi/a$. 
Given the above dimensional and R.G. arguments, $f(k^2/\L^2)$ is a 
dimensionless function independent on the scale $Q$, for large $Q$, and 
can thus be expressed in terms of a running coupling at the scale $k^2$. 
One obtains the Renormalon contribution by considering the high frequencies 
contribution to Eq.~(\ref{eq:Int1}), that is 
\beq\label{eq:Int2}
W^{\rm ren} =C\;\int^{Q^2}_{r\L^2}\; \frac{k^2\,dk^2}{Q^4}\; \as(k^2) \,,
\eeq
in which $f(k^2/\L^2)$ has been taken proportional to the perturbative 
running coupling (higher powers of the coupling simply result in subleading 
corrections to the formulas we will get this way). We now introduce the 
variable
\beq\label{eq:zvar}
z \equiv z_0\left(1-{\as(Q^2)}/{\as(k^2)}\right)
\,,
\;\;\;\;\;\;\;
z_0 \equiv \frac{1}{3b_0}\,,
\eeq
using which together with the two loop form for $\as(k^2)$ results in 
\beq\label{eq:borel}
W^{\rm ren}  = {\cal N} \int_{0}^{z_{0_-}} dz\;e^{-\be z}
\;(z_0-z)^{-1-\g} \,.
\eeq
In the last equation we have traded $\as(k^2)$ for the $\be$ coupling 
one is more familiar with on the lattice and introduced a couple of new 
symbols according to 
\beq
4\pi \as(Q^2)   \equiv  {6}/{\be} \,,
\;\;\;\;\;\;\;
\g \equiv 2\frac{b_1}{b_0^2} \,,
\;\;\;\;\;\;\;
0<z<z_{0_-}  \equiv  z_0(1-\as(Q^2)/\as(r\L^2)) \,.
\eeq
In the equations above $b_0$ and $b_1$ are the first and second 
coefficients of the perturbative $\be$--function (to fix normalisation: 
$b_0 = 11/(4\pi)^4$). 
$z_{0_-}$ is clearly reminiscent of the IR cutoff $r\L^2$ imposed in 
Eq.~(\ref{eq:Int2}) to avoid Landau pole once the perturbative coupling 
had to be plugged in. From Eq.~(\ref{eq:borel}) it is now easy to obtain 
a perturbative expansion
\beq\label{eq:cren}
W^{\rm ren}  = \sum_{\ell=1} \; \be^{-\ell} \;
\{c^{ren}_\ell + {\cal O}(e^{-z_0 \be})\} \,,
\;\;\;\;\;\;\;
c^{ren}_\ell = {\cal N'}\; \Gamma(\ell+\g)\;z_0^{-\ell}  \,.
\eeq
The factorial growth of the series is the Renormalon contribution, while 
the ${\cal O}(e^{-z_0 \be})$ is again coming from the IR cutoff $r\L^2$ 
in Eq.~(\ref{eq:Int2}). Naively one could have not imposed any IR cutoff 
before going from the ill-defined (because of the Landau pole) integral 
to the series. The factorial growth would have anyway asked for the 
${\cal O}(e^{-z_0 \be})$ contribution in order to obtain a resummation of 
the series. Actually the reader familiar with the subject has most probably 
well recognized in Eq.~(\ref{eq:borel}) a Borel representation. 

\subsection{The factorial growth on a finite lattice}\label{sec:fgfl}
Apart from the explicit reference to the lattice spacing in the UV cutoff 
we have till now adhered to a continuum notation, implicitly assuming some 
continuum scheme in which for example Eq.~(\ref{eq:Int1}) is defined. 
We now move to rewrite such a representation on a finite lattice. 
One can consider (notice that we now explicitly write the dependence on 
the number of lattice points $M$) 
\begin{equation}\label{eq:Wlat}
W^{\rm ren}(M) = C \; \int^{Q^2}_{Q_0^2(M)} \;
\frac{k^2\,dk^2}{Q^4} \; \as(s k^2) \,.
\end{equation}
The lower limit of integration is the explicit IR cutoff imposed by the 
finite extent of the lattice ($Q_0(M)=2\pi/Ma$), while the 
scale $s$ is in charge of matching from a continuum to a lattice 
scheme. In view of that and since rescaling $k^2 \rightarrow s k^2$ 
amounts to rescaling $\L \rightarrow s^{-1/2} \L$, it is likely to expect 
for $s$ a value $s=K^{-2}$ consistent (up to an eventual further change of 
scale) with $\L_{cont}=K\L_{latt}$ for some continuum scheme. By a change of 
variable in the integral it is easy to trade the rescaling of the argument 
of $\as$ for a rescale of $Q^2 \rightarrow s Q^2$ (and similarly for the 
lower limit of integration). 
Remember that in the previous section we obtained an expansion in 
the coupling defined at the scale $Q$: at one loop level a change of 
scale in the coupling is equivalent to a change of scheme, 
and this matches the present notation to the notation of 
\cite{8loop, L2}\footnote{For further details see later and the appendix.}. 
Given Eq.~(\ref{eq:Wlat}), one can of course perform the same 
change of variable defined in 
Eq.~(\ref{eq:zvar}) and obtain a new power expansion. In order not to have 
the main message of the procedure obscured by trivial algebra, we collect the 
details in an appendix. One is left with
\begin{equation}\label{eq:crenM}
W^{\rm ren}(M) = \sum_{\ell=1} \; \be^{-\ell} \, c^{ren}_\ell(M;s,C)
\end{equation}
The coefficients $c^{ren}_\ell(M;s,C)$ can be expressed in terms of 
incomplete $\Gamma$ functions (see the appendix). It is important to 
notice their dependences. They depend trivially ({\em i.e.} multiplicatively) 
on the overall constant $C$ which is present in Eq.~(\ref{eq:Wlat}) to 
account for everything which is subleading. They also depend on the scale 
$s$ which is in charge of the matching to the scheme in 
which Eq.~(\ref{eq:Int2}) best describes the expansion (remember that one 
does not know that {\em a priori}). Finally, they depend parametrically on 
the lattice size $M$. The attitude to take with respect to 
Eq.~(\ref{eq:crenM}) is now the same as in \cite{8loop,L2}, that is: can 
one determine values $C=\bar{C}$ and $s=\bar{s}$ by fitting the 
$c^{ren}_\ell(M;s,C)$ to the high orders $c^{(M)}_\ell$ ({\em i.e.} the 
coefficients actually computed on a given lattice size $M$)? 
A positive answer states that the perturbative 
expansion of the plaquette is actually described by the asymptotic 
dominance of the leading IR Renormalon. Needless to say, once the fit 
has been done and values $\bar{s}$ and $\bar{C}$ have been obtained, the 
$c^{ren}_\ell(M;\bar{s},\bar{C})$ contain in a sense all the information 
needed to test the IR Renormalon dominance. Given the data at our hand 
(coefficients $c^{(M)}_\ell$ for $\ell \leq \ell_{max}=10$ and $M=8,24$), 
one can perform a variety of checks. For example, one can fit $C=\bar{C}$ 
and $s=\bar{s}$ from data up to $\ell < \ell_{max}$ and then compare results 
at higher loops. As for finite size effects, one can fit the relevant 
parameters on a given lattice size $M$ (remember that the dependence on 
$M$ is a parametric one) and then compare 
$c^{ren}_\ell(M' \neq M;\bar{s},\bar{C})$ with the actual $c^{(M')}_\ell$. 
Such comparisons are what we actually proceed to do in the next section. \\
Before moving to the actual results, we think it is worthwhile to recall 
something about the crucial issue of finite size effects with respect to 
the dependence on both $M$ and $s$ of our formulae. The most straightforward 
way to obtain the IR Renormalon factorial growth goes through the most 
direct exploitation of Eq.~(\ref{eq:Int2}): one simply insert for 
$\as(k^2)$ the leading logs (one loop) formula. From this simple recipe 
one obtains the $\Gamma$ functions which are the signature of Renormalons. 
For a dimension $4$ condensate (as the observable at hand) order $l$ turns out 
to be proportional to (we write the explicit dependence on the IR cut--off) 
\beq
\int^{Q^2}_{Q_0^2}\; \frac{dk^2}{k^2}\; \frac{k^4}{Q^4}\; 
\ln^{\ell-1}(\frac{Q^2}{k^2})
\eeq
At this point a simple steepest descent argument points out that 
momenta $k^*$ that contribute the most to the $\ell$th order decrease 
exponentially with $\ell$ \cite{firstFV} 
\beq
k^* \sim e^{-(\ell-1)/2}
\eeq
and that's the reason for being {\em a priori} worried for a substantial 
taming of the factorial growth on a finite lattice: starting from a certain 
order momenta $k^*$ will fall outside the integration region. Now the point 
is that our formulae actually ask for the scale $s$ in front of the $k^2$ 
in the argument of the logarithm which results in 
\beq
k^* \sim s^{-1} e^{-(\ell-1)/2}.
\eeq
Being $s<<1$ this changes quite a lot the estimate of the dependence on 
$M$ of relatively high orders in Perturbation Theory. We will see that 
finite size effects are well in accord with our formulas and under control 
at the order we got.

\section{The Lattice $SU(3)$ plaquette to tenth order}
In Table~(1) we quote the coefficients of the expansion of 
the $SU(3)$ plaquette to tenth order on $8^4$ and $24^4$ lattices 
according to the notation of Eq.~(\ref{eq:clat}). 
The results were obtained on the {\em APE}--systems the Milan--Parma 
group were endowed with by {\em INFN}. Reaching ten order (up to the errors 
one can inspect from the table) required for the $24^4$ lattice about 
$2000$ hours on a {\em qh1} system 
(128 nodes, delivering a peak performance of $6.4$ 
Gflops). We stress that this system has got an expanded--memory; the $4$ 
MWords on each node allow space enough for a fairly big lattice to such an 
high order (we recall that in NSPT each field is replicated a number of 
time which is twice the order in $\alpha_s$, {\em i.e.} the order in $g$). 
Results for the $8^4$ lattice came out of about $4000$ hours on 
a {\em q4} system (32 nodes arranged in $4$ boards, on each of which a 
replica of the lattice was allocated; this system delivers a peak 
performance of $1.6$ Gflops). In considering these data and their statistical 
significance one should keep in mind that the signal to noise ratio is better 
for the bigger volume. 
\begin{table*}[t]
\caption{The first ten coefficients in the perturbative expansion of 
the Lattice $SU(3)$ plaquette.}
\begin{center}
\begin{tabular}{c|cc}
\hline
$n$          & $c^{(8)}_n$ & $c^{(24)}_n$ \\
\hline
$1$               & $1.9994(6) \cdot 10^0$ & $2.0000(4) \cdot 10^0$ \\
$2$               & $1.2206(16) \cdot 10^0$ & $1.2208(10) \cdot 10^0$ \\
$3$               & $2.9523(58) \cdot 10^0$ & $2.9621(48) \cdot 10^0$ \\
$4$               & $9.345(27) \cdot 10^0$ & $9.417(29) \cdot 10^0$ \\
$5$               & $3.397(14) \cdot 10^1$ & $3.439(20) \cdot 10^1$ \\
$6$               & $1.346(7) \cdot 10^2$ & $1.368(9)  \cdot 10^2$ \\
$7$               & $5.653(34) \cdot 10^2$ & $5.774(42)  \cdot 10^2$ \\
$8$               & $2.480(18) \cdot 10^3$ & $2.545(21)  \cdot 10^3$ \\
$9$               & $1.124(10) \cdot 10^4$ & $1.159(10)  \cdot 10^4$ \\
$10$              & $5.227(52) \cdot 10^4$ & $5.416(57)  \cdot 10^4$ \\
\hline
\end{tabular}\\[2pt]
\end{center}
\end{table*}
By inspecting the table one can make some first comments. First of all, we 
improved the errors with respect to \cite{8loop}. To improve errors on 
low loops would have been relatively easy, but since we were mainly interested 
in high loops, the simulation times (and therefore the statistical errors) 
were actually driven by the tenth order\footnote{The time spent on a single 
iteration basically scales as $(O^2-O)/2$, $O$ being the order in 
$\be^{-1/2}$. For more details on technical points concerning the 
application of NSPT to LGT see \cite{lgtNSPT}.}. 
Having said that, we decided we had the statistic significance we needed 
when finite size effects were disentangled from statistical errors. 
By inspecting the coefficients one can see that finite size effects between 
$8^4$ and $24^4$ are less than $1\%$ for the first four orders, less than 
$2\%$ for the first six, less than $3\%$ for the first eight and less 
than $4\%$ up to the highest order we got, {\em i.e.} tenth order. 
A proper infinite volume extrapolation is out of reach at this 
level, since one would need better statistic and of course more lattice 
sizes. Still, we will see that within the IR Renormalon dominance 
approach it is easy to estimate the correct order of magnitude of such 
an extrapolation. As we will see, up to tenth order one can infer a 
correction of the $24^4$ results with respect to infinite volume of less 
than $0.5\%$. \\
Having in mind all the points we have already made in the previous sections, 
we now approach the two basic questions that set the stage for this 
work, that is: Was the IR Renormalon growth correctly singled out in 
\cite{8loop,L2}? Are finite size effects under control? We will show 
that there is a positive answer for both questions. 

\subsection{The leading Renormalon behaviour}
In order to test the IR Renormalon dominance claimed in \cite{8loop,L2}, we 
now proceed to perform some consistency checks, just of the type we have 
already referred to in Section~(\ref{sec:fgfl}). These come out of three 
different fits which we now proceed to describe and which are in a sense 
a summary of a variety of checks we performed on our results. 

\begin{itemize}
\item
First of all we try to understand to which extent the ninth and tenth order 
(new) results could be inferred from the first eight (already computed) 
coefficients. In order to do that, we fit the formula for 
$c^{ren}_\ell(M;s,C)$ to the set 
$\{c^{(8)}_7, c^{(24)}_7, c^{(8)}_8, c^{(24)}_8\}$ (notice that while we want
to make contact with previous results, nevertheless we now take into account 
both lattice sizes). We will refer to this procedure as {\bf Fit 1}. 
As the result of this fit (and later of the following fits) we quote 
$s_1 = \sqrt{s}$ instead of $s$. This quantity is simply related to the 
matching between the $\L$ parameters between the continuum and lattice 
scheme one is quite familiar with
\beq\label{eq:similBen}
s_1 = \sqrt{s} = x \, \frac{\L_{latt}}{\L_{cont}}.
\eeq 

\begin{figure}[t]
\begin{center}
\mbox{\epsfig{figure=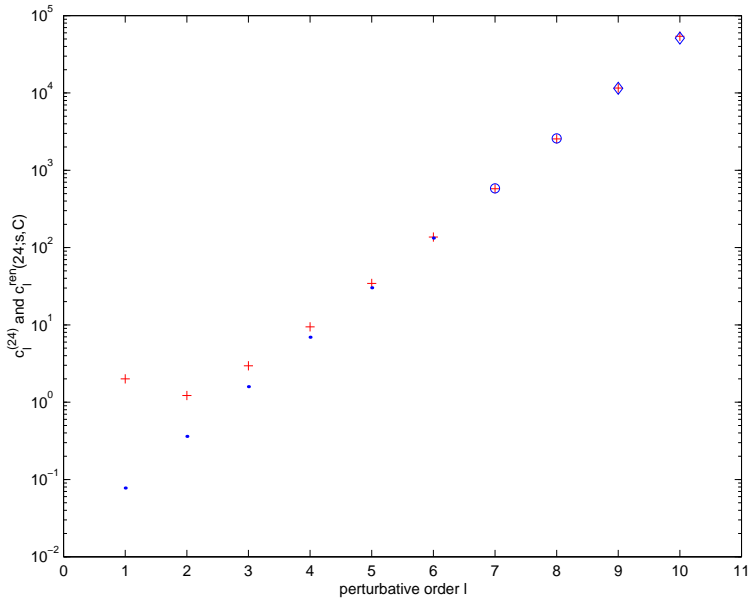,height=8.cm,width=10.cm}}
\caption
{The comparison between $c^{(24)}_\ell$ (red crosses) and 
$c^{ren}_\ell(24;s,C)$ as inferred from {\bf Fit1}. The latter are 
marked as circles for the orders that have been taken into account 
in the fit; they are instead marked as dots or diamonds for other points. 
In particular diamonds can be looked at as forecasts for the two highest 
orders computed.}
\end{center}
\end{figure}

\begin{table*}[b]
\caption{Perturbative coefficients $c^{ren}_\ell(M;s,C)$ as inferred 
from {\bf Fit1}. $\delta_M$ are per cent deviations from the actual 
values of $c^{(M)}_\ell$. $\delta_\infty$ are the residual per cent 
deviations of $24^4$ coefficients with respect to infinite volume.}
\begin{center}
\begin{tabular}{c||cc|cc||c}
\hline
          & $M=8$ & $\delta_8$ & $M=24$ & $\delta_{24}$ & $\delta_\infty$ \\
\hline
$c^{ren}_6$ & $1.30 \cdot 10^2$ & $- 3\%$ & $1.33 \cdot 10^2$ & $- 3\%$ & $ <0.1\%$ \\
$c^{ren}_7$ & $5.71 \cdot 10^2$ & $+ 1\%$ & $5.84 \cdot 10^2$ & $+ 1\%$ & $- 0.1\%$ \\
$c^{ren}_8$ & $2.51 \cdot 10^3$ & $+ 1\%$ & $2.58 \cdot 10^3$ & $+ 2\%$ & $- 0.1\%$ \\
$c^{ren}_9$ & $1.11 \cdot 10^4$ & $- 1\%$ & $1.15 \cdot 10^4$ & $- 1\%$ & $- 0.2\%$ \\
$c^{ren}_{10}$ & $4.90 \cdot 10^4$ & $- 7\%$ & $5.13 \cdot 10^4$ & $- 6\%$ & $- 0.3\%$ \\
\hline
\end{tabular}\\[2pt]
\end{center}
\end{table*}

In the previous formula the factor $x$ is a possible further change of 
scale (see later). We obtain a value for the the scale $s_1 = 0.0074$. 
In previous works by our group the fits were performed without referring 
to a change of scale in the integral. The matching from a continuum to the 
lattice scheme was obtained by a change of variable $\beta_{cont} = 
\beta_{latt} - r - \frac{r'}{\beta_{latt}}$. In terms of the notation 
of this work the result of the fit in \cite{L2} reads $s_1 = 0.0078$. 
The $5\%$ difference sets a rough order of magnitude for what one has to 
live with in this and the other comparisons that we make: one should always 
keep in mind that we are managing notations that differ by higher orders. 
Notice that this agreement is a first good message for our understanding 
of finite size effects based on our formula for $c^{ren}_\ell(M;s,C)$: 
since we now fit results both on the $8^4$ and on the $24^4$ lattices (while 
the results in \cite{L2} were based on $8^4$ alone) the parametric 
dependence on the lattice size $M$ is supposed to work pretty well (anyway 
see later for more definite statements). 
With respect to Eq~(\ref{eq:similBen}) Beneke \cite{beneke} made the point 
that the result $s_1 = 0.0078$ in \cite{L2} could for example be interpreted 
by matching to the $\MSbar$ scheme and taking into account a further change 
of scale (the factor $x$ in Eq~(\ref{eq:similBen})). The latter amounts to 
computing the $\MSbar$ coupling at a scale $0.706/a$ instead of $\pi/a$. 
With the second (standard) choice of scale one obtains the standard result 
$\L_{\MSbar} = 28.8 \, \L_{latt}$. $s_1 = 0.0078$ could in turn be 
interpreted as $0.706/\pi \times 1/28.8 \, = 0.0078$, while the result 
$s_1 = 0.0074$ could be read $0.666/\pi \times 1/28.8 \, = 0.0074$. One should 
not take all these considerations too seriously, {\em e.g.} there is no 
compelling commitment to the $\MSbar$ scheme (once again, remember that 
one does not know which is the scheme in which our theoretical prejudice 
works the best). Still, it is reassuring that 
the numbers one gets are absolutely sensible as for their order of 
magnitude. Having made contact with our previous results, we now proceed 
to make contact with what we are mainly interested in, that is the new 
(ninth and tenth) orders. One can get a first glance at this by 
looking at Fig.~(1) in which we compare the computed 
$c^{(24)}_\ell$ with the $c^{ren}_\ell(24;0.0074^2,C^{(1)})$ (we will not 
quote the values obtained for the overall constant $C$ since they are not 
too enlightening). Notice that different symbols refer to different orders 
as far as the $c^{ren}_\ell(24;0.0074^2,C^{(1)})$ are concerned: we want to 
emphasize the difference between the orders that are actually included 
in the fit and those that 
are looked at as {\em forecasts} to assess how well the asymptotic 
behaviour has been singled out. A very good agreement is manifest, which is 
actually magnified by the logarithmic scale. A careful inspection 
of the numbers themselves can be got from Table~(2) in 
which we quote the values of the last five coefficients as inferred from 
{\bf Fit 1} on both lattice sizes. A first point to be made has to do with 
finite size effects. As we have already mentioned, a fit to data on 
both sizes is absolutely sensible and the deviations of the inferred 
coefficients from the actual ones are much the same on $8^4$ and on $24^4$. 
Notice also that the residual finite size effects between $24^4$ and 
infinite volume turn out to be negligible. As for the crucial issue of 
how well {\bf Fit 1} can single out the asymptotic behaviour we notice 
that the ninth coefficient is forecast with a precision of $1\%$, while 
a $6$--$7\%$ deviation is there for the tenth order. Notice also that, 
while orders seven and eight are kept into account in the fit, the 
agreement for orders less than seven results in a sensible shape for an 
asymptotic behaviour (by the way, the results are pretty stable to the 
inclusion of order six in the fit). This is the right time 
to make a point that specifies a statement that we have already made, 
{\em i.e.} that in inspecting the forecasts of our fits one should keep 
in mind that we are managing expressions that are there up to higher orders. 
This is of course not the end of the story. 
An equally important indetermination is coming from the fact 
that only {\em a posteriori} one can understand {\em how asymptotic} the 
expansion was at the highest order which has been taken into account in 
the fit. Roughly speaking, by fitting only the expansion for one quantity 
there is no obvious way to prevent the fit from pretending to have already 
reached the actual asymptotic regime. 

\item
With respect to the last point we made there is a nice way to improve. 
We exploit it in what we call 
{\bf Fit 2}. This time we make use of the fact that in \cite{8loop} results 
for the first eight orders were computed not only for the $1\times1$ 
Wilson loop (the basic plaquette), but also for the $2\times2$. 

\begin{figure}[htb]
\begin{center}
\mbox{\epsfig{figure=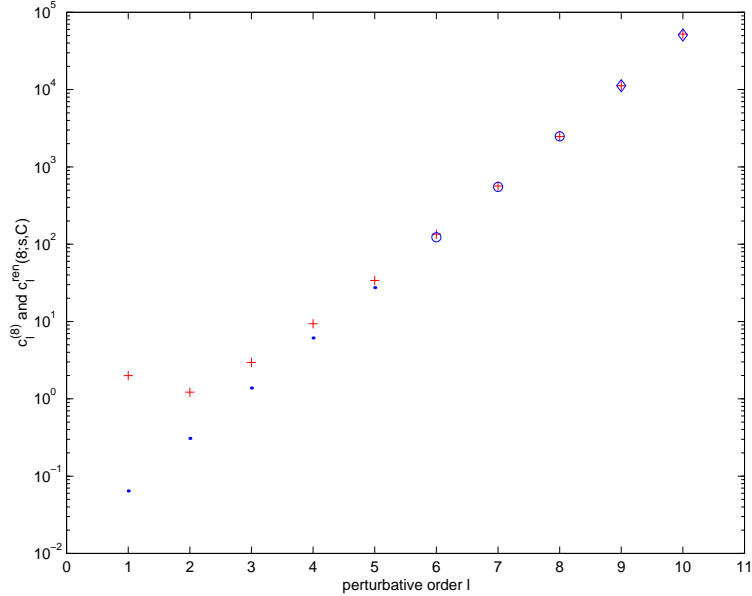,height=8.cm,width=10.cm}}
\caption
{The comparison between $c^{(8)}_\ell$ (red crosses) and 
$c^{ren}_\ell(8;s,C)$ as inferred from {\bf Fit2}. Same notations as in 
Fig.~(1).}
\end{center}
\end{figure}

\begin{table*}[b]
\caption{Perturbative coefficients $c^{ren}_\ell(M;s,C)$ as inferred 
from {\bf Fit2}. Same notations as in Tab.~(2).}
\begin{center}
\begin{tabular}{c||cc|cc||c}
\hline
          & $M=8$ & $\delta_8$ & $M=24$ & $\delta_{24}$ & $\delta_\infty$ \\
\hline
$c^{ren}_6$ & $1.23 \cdot 10^2$ & $- 10\%$ & $1.25 \cdot 10^2$ & $- 10\%$ & $ <0.1\%$ \\
$c^{ren}_7$ & $5.51 \cdot 10^2$ & $- 2\%$ & $5.64 \cdot 10^2$ & $- 2\%$ & $- 0.1\%$ \\
$c^{ren}_8$ & $2.48 \cdot 10^3$ & $+ 0.2\%$ & $2.55 \cdot 10^3$ & $+ 0.3\%$ & $- 0.1\%$ \\
$c^{ren}_9$ & $1.12 \cdot 10^4$ & $- 0.2\%$ & $1.16 \cdot 10^4$ & $+ 0.2\%$ & $- 0.2\%$ \\
$c^{ren}_{10}$ & $5.09 \cdot 10^4$ & $- 3\%$ & $5.31 \cdot 10^4$ & $- 2\%$ & $- 0.3\%$ \\
\hline
\end{tabular}\\[2pt]
\end{center}
\end{table*}

Now, the IR Renormalon dominance is supposed to force a {\em universal} 
asymptotic behaviour. That's why we now fit the formula for 
$c^{ren}_\ell(M;s,C^{(i)})$ to the set 
$\{c^{(8)}_6, c^{(8)}_7, c^{(8)}_8, \\
c^{(8,2\times2)}_6, 
c^{(8,2\times2)}_7, c^{(8,2\times2)}_8\}$. The $c^{(8,2\times2)}_\ell$ 
are taken from \cite{8loop}, while of course this time a couple of 
different overall constant $C^{(i)}$ have to be fitted, corresponding to 
$i=1\times1$ and $i=2\times2$. Notice that this time, as an extra 
test of the control on finite size effects, we make use only of 
$8^4$ results for the basic plaquette. The value fitted for the scale is now 
$s_1 = 0.0065$. While we are still stopping at eight order in what 
we take into account for the fit, the curvature in the growth of the 
$2\times2$ coefficients is opposite to that of $1\times1$ and this corrects 
a bit with respect to {\bf Fit 1}. As for the $12\%$ change in the scale, 
much the same that has already been told holds: while there is in a sense 
extra information on what has to be understood as asymptotic, it is 
nevertheless reassuring that the change in the coefficients is quite 
smooth and absolutely sensible as we are managing asymptotic behaviours. 
Not surprisingly, one can inspect from 
Fig.~(2) (this time we plot the comparison for coefficients 
on the $8^4$ lattice) and (even better) from Table~(3) 
that we have fairly improved the asymptotic behaviour. In particular, 
deviations for the forecasts on order ten are reduced to $1$--$2\%$, while 
other nice features like the consistency of finite size effects are still 
there. Also the residual finite size effects dependency turns out to be 
just the same (negligible) order. 

\item
One can of course devise a fit in which all the information at hand is 
taken into account. This is what we proceed to do in what we refer to as 
{\bf Fit 3}. This time we fit the formula for $c^{ren}_\ell(M;s,C^{(i)})$ 
to the set $\{c^{(8)}_6, c^{(8)}_7, c^{(8)}_8, c^{(8)}_9, c^{(8)}_{10}, 
c^{(24)}_6, c^{(24)}_7, c^{(24)}_8, c^{(24)}_9, c^{(24)}_{10}, 
c^{(8,2\times2)}_6, \\
c^{(8,2\times2)}_7, c^{(8,2\times2)}_8\}$. 
\begin{figure}[t]
\begin{center}
\mbox{\epsfig{figure=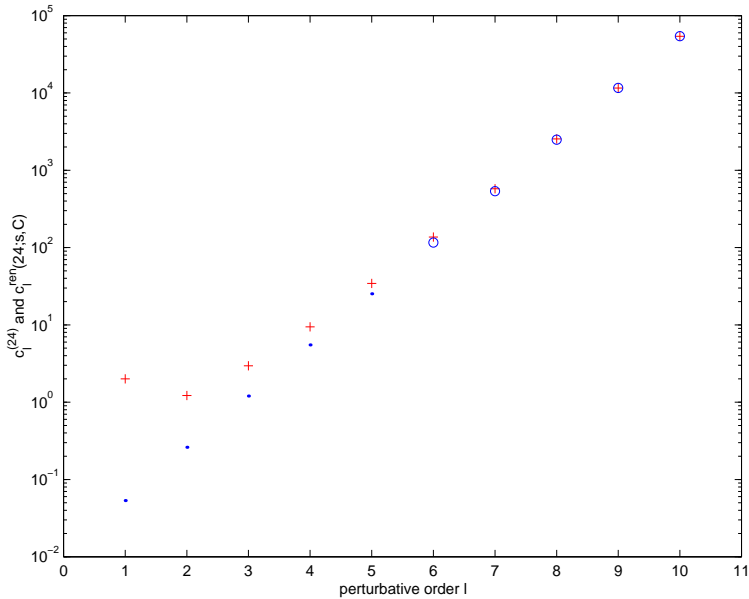,height=8.cm,width=10.cm}}
\caption
{The comparison between $c^{(24)}_\ell$ (red crosses) and 
$c^{ren}_\ell(24;s,C)$ as inferred from {\bf Fit3}. Same notations as in 
Fig.~(1). Notice that this time there are no diamonds.}
\end{center}
\end{figure}
\begin{table*}[b]
\caption{Perturbative coefficients $c^{ren}_\ell(M;s,C)$ as inferred 
from {\bf Fit3}. Same notations as in Tab.~(2).}
\begin{center}
\begin{tabular}{c||cc|cc||c}
\hline
          & $M=8$ & $\delta_8$ & $M=24$ & $\delta_{24}$ & $\delta_\infty$ \\
\hline
$c^{ren}_6$ & $1.14 \cdot 10^2$ & $- 17\%$ & $1.16 \cdot 10^2$ & $- 17\%$ & $ <0.1\%$ \\
$c^{ren}_7$ & $5.26 \cdot 10^2$ & $- 7\%$ & $5.37 \cdot 10^2$ & $- 7\%$ & $- 0.1\%$ \\
$c^{ren}_8$ & $2.43 \cdot 10^3$ & $- 2\%$ & $2.49 \cdot 10^3$ & $- 2\%$ & $- 0.1\%$ \\
$c^{ren}_9$ & $1.12 \cdot 10^4$ & $+ 0.2\%$ & $1.16 \cdot 10^4$ & $+ 0.4\%$ & $- 0.2\%$ \\
$c^{ren}_{10}$ & $5.21 \cdot 10^4$ & $< 0.1\%$ & $5.43 \cdot 10^4$ & $+ 0.5\%$ & $- 0.3\%$ \\
\hline
\end{tabular}\\[2pt]
\end{center}
\end{table*}
We are again changing the players on the ground by including a wide range 
of orders on both lattice sizes, together with information from $2\times2$ 
Wilson Loop as well. The resulting scale is now $s_1=0.0056$, with a 
change with respect to {\bf Fit 2} which is again the same order of 
magnitude as that obtained in going from {\bf Fit 1} to {\bf Fit 2}. 
The result is remarkably stable with respect to keeping 
into account only the last three orders for the basic plaquette (as for 
the $2\times2$ Wilson Loop one can at the same time keep into account 
only the last order available, that is order eight). Results of the fit 
are plotted in Fig.~(3) for the $24^4$ lattice and 
summarized in Table~(4). Of course, this time 
it does not make any sense to compare with {\bf Fit 1} and {\bf Fit 2} 
as for the accuracy with which high orders are described. 
\end{itemize}
In the end, what one can 
state is that going from {\bf Fit 1} to {\bf Fit 3} one is not going to 
jeopardize the overall picture: the transition in describing the growth 
of the coefficients is quite smooth, consistent with the asymptotic 
behaviour being better and better described. Again, one is validating 
the finite size effects as embedded in our formulae.

\subsection{Finite size effects}
From all the previous arguments the impact of finite size effects has 
already been widely discussed. Still, we think it is worthwhile to 
present the tests of finite size effects also in a graphical format. 
In Fig.~(4) we depict something that has to do with 
finite size effects as resulting from the results of {\bf Fit 2}. 
\begin{figure}[htb]
\begin{center}
\mbox{\epsfig{figure=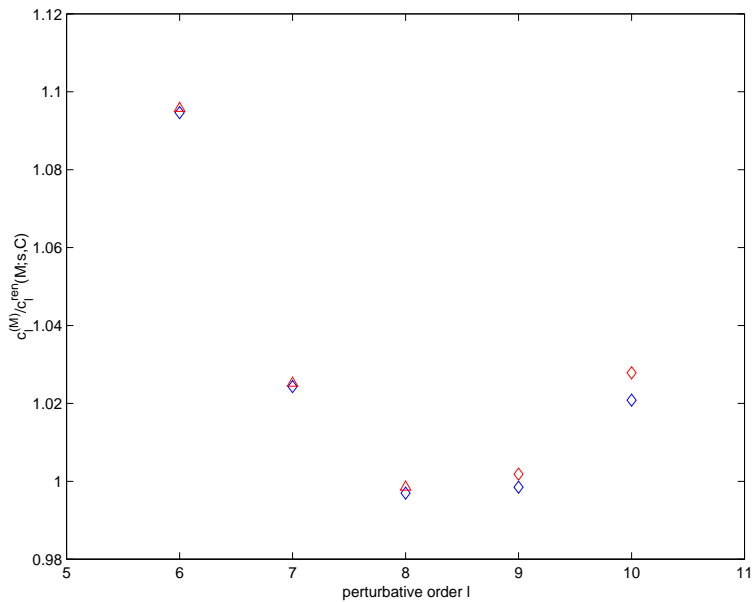,height=8.cm,width=10.cm}}
\caption
{The ratios ${c^{(8)}_\ell}/{c^{ren}_\ell(8;s,C)}$ (red triangles pointing 
up and red diamonds) and ${c^{(24)}_\ell}/{c^{ren}_\ell(24;s,C)}$ 
(blue diamonds) as inferred from the results of {\bf Fit 2}.}
\end{center}
\end{figure}
To be definite, we plot the ratios 
${c^{(8)}_\ell}/{c^{ren}_\ell(8;0.0065^2,C^{(1\times1)})}$ and 
${c^{(24)}_\ell}/{c^{ren}_\ell(24;0.0065^2,C^{(1\times1)})}$. 
One can directly see the conclusion that has already been drawn in the main 
discussion of {\bf Fit 2}. While only the $8^4$ lattice has been taken into 
account in the fit, the points for both lattices fall on top of each other, 
{\em i.e.} from a fit to $8^4$ and the dependence on $M$ embedded in our 
formulae we describe basically with the same accuracy the coefficients of 
the $24^4$ lattice as well. Something similar is shown in 
Fig.~(5) which is the same plot for the equivalent of 
{\bf Fit 3} in which only the $8^4$ lattice has been taken into account. 
\begin{figure}[htb]
\begin{center}
\mbox{\epsfig{figure=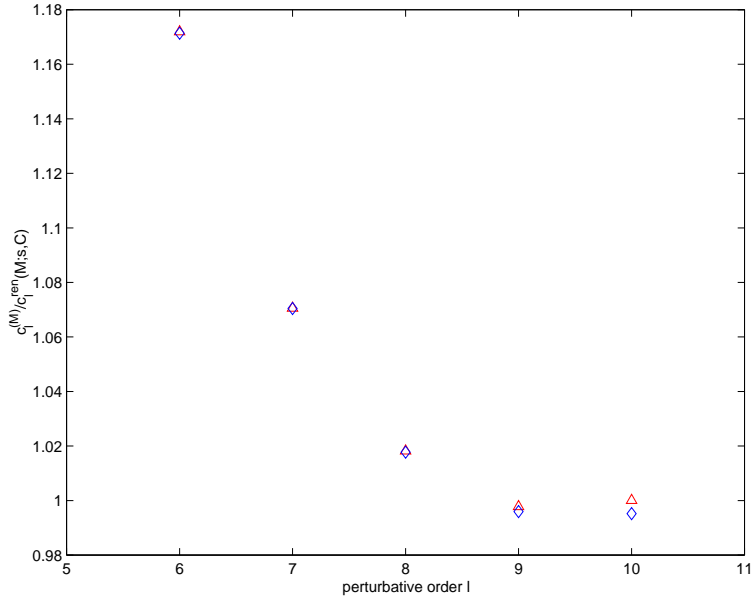,height=8.cm,width=10.cm}}
\caption
{Same quantities as in Fig.~(4) as obtained by the results of the 
equivalent of {\bf Fit 3} in which only $8^4$ lattice has been taken into 
account. Quantities for $8^4$ are plotted as red triangles pointing up 
while quantities for $24^4$ are plotted as blue diamonds.}
\end{center}
\end{figure}

\section{Conclusions and perspectives}
We computed the perturbative expansion of the basic $SU(3)$ plaquette on 
both a $8^4$ and a $24^4$ lattices to $\be^{-10}$ order. 
We are now in a position to strengthen 
the claim in \cite{8loop,L2}, {\em i.e.} the coefficients growth is 
consistent with both the leading IR Renormalon dominance and with finite 
size effects on top of that. The latter does not exceed $4\%$ at the order 
we got and the residual finite size effects on the $24^4$ results are 
no more than half a percent, again at tenth order. Having verified that 
our previous analysis in \cite{L2} had neither failed in the extraction 
of the leading asymptotic behaviour nor underestimated finite size effects 
on the perturbative coefficients, these results make even more important 
to refine the analysis contained in the same paper as for the extraction 
of a quadratic contribution to the lattice representative for the Gluon 
Condensate. This refinement will now benefit by a control on 
finite size effects. This is what we plan to do in a near future 
\cite{testL2}, maybe going even through a further refinement on errors 
in the perturbative coefficients.

\vskip 1cm
\noindent
\section*{Acknowledgments}
\par\noindent
The authors are grateful to G. Burgio for many interesting discussions 
and to E. Onofri and G. Marchesini for their constant interest and 
encouragement. F.~D.R. acknowledges support from both Italian MURST 
under contract 9702213582 and from I.N.F.N. under {\sl i.s. PR11}. L.~S. 
acknowledges support from EU Grant EBR-FMRX-CT97-0122.

\section*{Appendix}
\par\noindent
We now sketch the steps to go from Eq.~(\ref{eq:Wlat}) to 
Eq.~(\ref{eq:crenM}). As we have already said in Sec.~(\ref{sec:fgfl}) 
one way to treat the dependence on $s$ is to rescale the integration 
variable (define $t^2 \equiv s k^2$). In this way one ends up with the 
rescaling $Q^2 \rightarrow s Q^2$. This in turns means 
$\be \rightarrow \be(s Q^2)$ which makes contact with the formulation 
of \cite{8loop,L2} (again, as already said): $\be \rightarrow \be 
- r$ (at this level one obtains a one loop formula). We chose another way 
to proceed which manipulates the integrand in another way. By going through 
the change of variable of Eq.~(\ref{eq:zvar}) one obtains 
(up to overall constants) 
\beqn
\int_{0}^{z_{ir}(M)} dz\;e^{-\be z} \;(z_0-z)^{-1-\g} 
\frac{1}{1+6 b_0 \frac{z_0}{z_0-z}\frac{1}{\be} \ln s}\,, \non
\eeqn
$z_{ir}(M)$ being the value for $z$ pertaining to the IR cut--off $Q_0(M)$. 
The last factor is simply $\frac{\as(s k^2)}{\as(k^2)}$. By expanding 
the latter in a geometric series (we are aiming at an expansion 
in $\be^{-1}$) and by performing a change of variable we end up with
\beqn
\be^\gamma e^{-\be z_0} (-1)^{-1-\gamma} 
\sum_n (6 b_0 z_0 \ln s)^n \int_{-\be z_0}^{\be z_{ir}(M) - \be z_0} 
e^{-z} z^{-1-\gamma-n} dz \non
\eeqn
Note that the effect of the scale $s$ propagates to high orders much 
the same way the leading logs expansion for $\as$ is responsable for the 
Renormalon growth, \ie via a geometric series. 
The last input is now the asymptotic expansion for incomplete $\Gamma$ 
functions (which is useful once one splits the previous integral in terms 
of a sum of incomplete $\Gamma$ functions)
\beqn
\int_z^\infty dt \, e^{-t} \, t^{a-1} \approx z^{a-1} e^{-z} 
\sum_{k\geq 0} \frac{\Gamma(k+1-a)}{\Gamma(1-a)} (-z)^{-k} \non
\eeqn
Once this expansion (valid for $|z|\rightarrow \infty$ 
in $|\arg z|<3 \pi/2$) is plugged in, one proceeds to the final power 
expansion in $\be^{-1}$ (which is easy to manage for example in 
$Mathematica^\copyright$).


\end{document}